\documentclass[iop]{emulateapj}

\usepackage{graphicx}
\usepackage{amsmath}
\usepackage{kantlipsum}
\usepackage{xparse}
\usepackage{enumerate}
\usepackage[version=3]{mhchem}
\usepackage{amsmath,array}
\usepackage[T1]{fontenc}
\NewDocumentCommand{\codeword}{v}{%
\texttt{\textcolor{blue}{#1}}%
}
\begin{document}

\title{Coupled thermal and compositional evolution of photo evaporating planet envelopes}

\author{Isaac Malsky\altaffilmark{1, 2}, Leslie A. Rogers\altaffilmark{1}}
\affil{Astronomy Department, University of Chicago, Chicago, IL 60637}
\affil{Astronomy Department, The University of Michigan, Ann Arbor, MI, 48109}
\begin{abstract}
Photo-evaporative mass loss sculpts the atmospheric evolution of tightly-orbiting sub-Neptune-mass exoplanets. To date, models of the mass loss from warm Neptunes have assumed that the atmospheric abundances remain constant throughout the planet's evolution.
However, the cumulative effects of billions of years of escape modulated by diffusive separation and preferential loss of hydrogen can lead to planetary envelopes that are enhanced in helium and metals relative to hydrogen \citep{2015ApJ...807....8H}.
We have performed the first self-consistent calculations of the coupled thermal, mass-loss, and compositional evolution of hydrogen-helium envelopes surrounding sub-Neptune mass planets. 
We extended the MESA (\textit{Modules for Experiments in Stellar Astrophysics}) stellar evolution code to model the evolving envelope abundances of photo-evaporating planets.
We find that GJ~436b, the planet that originally inspired \citet{2015ApJ...807....8H} to propose the formation of helium enhanced planetary atmospheres, requires a primordial envelope that is too massive to become helium enhanced.
Nonetheless, we show that helium enhancement is possible for planets with masses similar to GJ 436b after only several Gyr of mass loss. These planets have $R_p\lesssim 3.00~R_\oplus$, initial $f_{\rm env} < 0.5\%$, irradiation flux $\sim$10$^1$--10$^3$ times that of Earth, and obtain final helium fractions in excess of Y=0.40 in our models. The results of preferential envelope loss may have observable consequences on mass-radius relations and atmospheric spectra for sub-Neptune populations.
\end{abstract}

\section{Introduction}\label{sec:Intro}
No solar system analogues exist for the thousands of tightly orbiting sub-Neptune mass exoplanets discovered, inspiring recent work about the nature of these planets \citep[e.g.,][]{Howard2012, Kepler1, Kepler2}. In recent years, observational work combined with theoretical formation and evolution models have shown the incredible diversity possible in this planetary population. Mass---radius ($M_p$---$R_p$) measurements and statistical analyses provide evidence that many of these planets have an envelope of light gases, increasing their observed radii \citep{Marcy2014, RockeyCorePlus}. A large portion of these planets are tightly orbiting (with orbits interior to $0.25$~AU) and have masses between 1 and 25~$M_{\oplus}$ \citep{Rowe, BoruckiEt2011ApJ, 2016AJ....152..187M}. Throughout this work, we refer to warm Neptunes and sub-Neptunes as planets with masses below 25.0 $M_\oplus$ and orbits interior to 1.0 AU.

A key to understanding exoplanet demographics lies in analyzing the interplay between a planet's H/He envelope and incident irradiation in the evolution of these tightly orbiting sub-Neptune mass planets. Evolution and structure models have constrained the $M_p$---$R_p$ relations for low density exoplanets, and shown that the presence of a volatile envelope can greatly inflate a planet's radius \citep{RogersEt2011ApJ}. \cite{2014ApJ...792....1L} showed that changing the H/He envelope mass fraction of a planet has a dramatic effect on planetary radius, subsuming the smaller effects due to incident flux and planet age. Evaporative mass loss depletes the gas envelopes of sub-Neptune-size planets, significantly decreasing their radii \citep[e.g.,][]{2012MNRAS.425.2931O, Chen&Rogers2016ApJ, 2013ApJ...776....2L, Howe_2014,
2010A&A...516A..20V}.

Theoretical analyses predict that photo-evaporation can shape the evolution of highly irradiated sub-Neptunes, bifurcating the population based on envelope retention (\citealt{2013ApJ...776....2L, 0004-637X-775-2-105, 2018arXiv180707609O}). Under the most intense irradiation, atmospheric ``boil off' creates planets with radii similar to that of their heavy element (rocky or icy) cores. Under less extreme envelope erosion, planets retain a volatile envelope, and have subsequently larger radii. \cite{2017ApJ...847...29O} predict that the photo-evaporative process leaves a H/He envelope (for planets with final radii $\sim$ 2.6 $R_\oplus$), or strips planets of their entire envelope (with radii of $\sim$ 1.3 $R_\oplus$) over the course of 100 Myrs.

Observational surveys have shown the existence of a bimodal radius distribution of small planets similar to that predicted by the photo-evaporation valley, with occurrence rates peaked at $R_{p} < 1.5~R_{\oplus}$ and $2.0~R_{\oplus} < R_{p} < 3.0~R_{\oplus}$ \citep{2017AJ....154..109F}.  \cite{2016NatCo...711201L} find no exoplanets with radii between 2.2 - 3.8 $R_\oplus$ which have incident flux rates greater than 650 times that of Earth. Furthermore, few sub-Neptune planets have been discovered with orbital periods below 2-4 days, indicating a desert of ultra-irradiated sub-Neptune mass exoplanets \citep{2016A&A...589A..75M, 2011A&A...528A...2B, 2013ApJ...763...12B}.

Theories of how atmospheric erosion is shaping exoplanet populations are becoming increasingly important to interpret observational results.
To date, most mini-Neptune interior structure models have assumed solar H/He mass ratios \citep[e.g.,][]{Chen&Rogers2016ApJ, 2014ApJ...792....1L, 2015ApJ...808..150H, ValenciaEt2007bApJ}. Additionally, models of photo-evaporation \citep[e.g.,][]{EUVFlux, Lopez1, 0004-637X-775-2-105, 2017ApJ...847...29O, Lehmer_2017} have assumed that envelope composition stays constant over time, and that planets lose hydrogen, helium, and metals in the same proportions as they are present in their envelopes.

Preferential loss of light gases can cause planets to become enhanced in helium and metals relative to hydrogen over the course of billions of years. \cite{2015ApJ...807....8H} have proposed that H/He fractionation and envelope mass loss creates planets with orders-of-magnitude reductions in atmospheric hydrogen. Stellar EUV radiation causes hydrodynamic outflow of atmospheric gases, with mass flux proportional to the mean molecular mass of each gas. For helium enhancement, planets must have relatively small initial envelope mass fractions ($\sim 10^{-3}$). Basing their model off previous interior structure models \citep{2010A&A...523A..26N}, \cite{2015ApJ...807....8H} proposed that helium enhancement could explain the lack of CH$_4$ in GJ 436b's emission spectrum. Although \cite{2015ApJ...807....8H} modeled time-varying composition of planet envelopes, they did not couple this to a model of the interior structure of the planet and its thermal and radius evolution.

In this paper, we develop novel methods to model coupled thermal and compositional evolution in warm Neptunes. First, we analyze the effect of varying envelope helium fraction of the interior structure and radius of exoplanet populations. We create $M_p$---$R_p$ relations with a range of initial envelope mass fractions and helium fractions. Second, we develop the first simulations of the coupled thermal, mass-loss, and envelope composition evolution of exoplanets. Third, we test the mechanism of mass loss proposed by \cite{2015ApJ...807....8H} with GJ 436b to determine whether it is a candidate for helium enhancement.

We present our methodology for creating planetary models with MESA in \S~\ref{sec:methods}. In \S~\ref{sec:results}, we describe the results of the simulations of atmospheric structure and envelope mass loss. In \S~\ref{sec:disc}, we discuss the effects of coupled thermal, mass-loss, and envelope composition evolution. Finally, results and conclusions are in \S~\ref{sec:conc}.

\section{Methods}\label{sec:methods}
We use the Modules for Experiments in Stellar Astrophysics (MESA, \citealt{2011ApJS..192....3P, 2013ApJS..208....4P, 2015ApJS..220...15P, PaxtonEt2018ApJS}), an open source Fortran library for stellar evolution, to model the evolutionary pathways of exoplanets. We specifically use MESA version 10,398. Multiple studies have already applied MESA to exoplanets with H/He-dominated envelopes \citep[e.g.,][]{2015ApJ...813..101V, 0004-637X-817-2-107, 0004-637X-821-1-26,  Chen&Rogers2016ApJ, 2017ApJ...835..145J}, but all have so far assumed that the composition of the planet's envelope (specifically  the hydrogen mass fraction, $X$, helium mass fraction $Y$, and heavy element mass fraction, $Z$) remains constant throughout time. 

We extend MESA to simulate the compositional evolution of Neptune-size and sub-Neptune-size planet envelopes undergoing atmospheric escape. Our models build upon the MESA test suites \verb|irradiated_planet| and \verb|make_planets|, and the previous sub-Neptune planet modeling approach from \citet{Chen&Rogers2016ApJ}. We have further standardized the procedures for creating initial planet models with varied envelope compositions (\S~\ref{sec:ic}), updated the treatment of the irradiation on the planet to include physical opacities (\S~\ref{sec:opacity}), and implemented the mass loss prescription from \citet{2015ApJ...807....8H} (\S~\ref{sec:H/He Mass Loss}). We elaborate upon each of these new developments below.

\subsection{Creating Planetary Models with MESA}\label{sec:ic}
Our simulations are split into two distinct stages. First, we bring an initial planetary model through a series of iterative steps that gradually relax its properties until the desired initial conditions (planet mass, $M_p$, core mass, $M_{\rm core}$, envelope composition, $X$, $Y$, $Z$, initial entropy, $S$, and irradiation flux, $F_\star$) are attained. MESA natively models stars; creating starting models for Earth-mass-scale planets is non-trivial. Once the initial model is brought to the desired starting conditions, the planet is evolved for several billions of years. It is in this second stage that the adjustments to the planet model from one time step to the next encapsulate the physics of planet thermal evolution and mass loss. 

Though we base our procedure for creating initial planet models on that outlined in \cite{Chen&Rogers2016ApJ}, we have made substantial adjustments both to improve stability and to model planets across a wider parameter space. Compared to \cite{Chen&Rogers2016ApJ}, we iterate back and forth through several rounds of planet mass reduction and heavy-element core insertion, until the desired $M_p$ and $M_{\rm core}$ are reached. We also include functionality to adjust the planet envelope composition and create planet starting models with non-solar atmospheric compositions.

Our procedure to create of each initial planet model follows the steps below. 

\begin{enumerate}
\item First we load an initial planet profile -- \verb|0.001Msun.mod| from \verb|very_low_mass_grey_models|.
\item The planet is reduced in mass to 166.50 $M_\oplus$ using \verb|relax_mass|.
\item Next, a rocky core is inserted with a mass of 0.67 $M_\oplus$ using $\verb|relax_core|$. As in \cite{Chen&Rogers2016ApJ}, we use the core mass-radius relations from \cite{RogersEt2011ApJ} to specify the mean density of the heavy-element core. The model core from \verb|relax_core| is inert. However, we do make use of the time-varying core luminosity routine from \cite{Chen&Rogers2016ApJ} in the final evolution stage.
\item We then reduce planet mass to 30 times its final mass. Adjusting planet mass and core mass in stages increases model stability.
\item Next, we relax the rocky core to 10\% of the final core mass using \verb|relax_core|. 
\item At this point, we relax $X$, $Y$ and $Z$ to the initial envelope composition desired using \verb|relax_initial_Y| and \verb|relax_initial_Z|. 
\item Next, the model evolves in isolation for $10^6$ years to stabilize before continuing with the ensuing parameter adjustments.
\item We then further reduce the planet mass to its final value of between 2.0 and 25.0 $M_\oplus$.
\item We increase the core mass to the ultimate value of $M_{\rm core}$ desired (taking care to adjust the core density, as appropriate).
\item The next two steps aim to standardize the initial thermal state of the planets by setting the initial interior entropy, $S$. If the planet's central entropy at this point is lower than the target value, we add an artificial luminosity (using \verb|relax_initial_L_center|) to the core to re-inflate the planet until the target $S$ is surpassed. If the planet's central entropy is already higher than the target value, this step is skipped. 
\item The artificial core luminosity is removed (if present), and the planet is allowed to cool until the desired initial entropy is reached (specified through \verb|center_entropy_lower_limit|). \label{pre-irrad}
\item At this point, we iteratively solve  for the appropriate value for \verb|column_depth_for_irradiation| as described in \S~\ref{sec:opacity}, and then relax the stellar irradiation incident on the planet to the desired value, $F$, using \verb|relax_irradiation|. 
\item Finally, the planet age is reset to zero, and we evolve the model for a short time (on the order of a few Myr) without mass loss.
\end{enumerate}

During the final step, we adopt a value for the heat capacity of the rocky core using the same routine as in \cite{Chen&Rogers2016ApJ}, and set $c_v$ to 1.0 J K$^{-1}$ g$^{-1}$ \citep{1995ApJ...450..463G}. A routine for adding the luminosity from the heavy element interior to the base of the envelope is implemented, as in \cite{Chen&Rogers2016ApJ}, including contributions from both the cooling of the core and radioactive heating.

Once the steps above are complete, the initial model is ready for simulating the simultaneous thermal, mass loss, and envelope composition evolution of the planet. Upon publication, scripts and inlists for creating initial planet models following our recipe will be made available on the MESA marketplace. 

\subsection{Irradiated Atmospheric Boundary Condition}\label{sec:opacity}
We use MESA's built-in $F_\star-\Sigma_\star$ surface heating \citep{2013ApJS..208....4P} to account for the irradiation incident on the planet from its host star. This heating function deposits the specified irradiation flux, $F_\star$ (corresponding to inlist option \verb|irradiation_flux|), in the outer layers of the planet's envelope down to the specified column depth, $\Sigma_\star$ (corresponding to inlist option \verb|column_depth_for_irradiation|). Note, we are taking a different approach to the atmospheric boundary conditions for irradiated planets than \cite{Chen&Rogers2016ApJ}, who used a modified version of the MESA \verb|grey_irradiated| atmospheric boundary condition. The dayside flux ($F_\star$) absorbed by the planet is determined by the stellar effective temperature, $T_{\rm eff \star}$, the planet's orbital separation, $d$, and the planet's Bond albedo, $A$, 

\begin{equation}
F_\star = \sigma T_{\rm eff \star}^4\left(\frac{R_\odot}{d}\right)^2\left(1-A\right).
\label{Solar_Flux}
\end{equation}

\noindent The planet's equilibrium temperature, $T_{\rm eq}$, is related to $F_\star$ via,

\begin{equation}
T_{\rm eq} = \left(\frac{F_\star }{4~\sigma}\right)^\frac{1}{4}.
\label{Planet_teff}
\end{equation}

The extent to which stellar irradiation penetrates the envelope is governed by the atmospheric opacity, $\kappa_v$, to the incoming  starlight, 
\begin{equation}
\Sigma_{*} =  2 / \kappa_{\nu}.
\label{ColumnDepth}
\end{equation}
\noindent We use gaseous mean opacities to incident stellar radiation from  \citet{2014ApJS..214...25F} to determine $\Sigma_\star$, as elaborated below. This goes a step beyond past studies that have specified a constant (semi-arbitrary) value of $\Sigma_\star$ \citep[e.g.,][]{2015ApJ...813..101V}. Using tabulated opacities to determine $\Sigma_\star$ allows better modeling of how stellar optical radiation is absorbed in a planet's atmosphere.

To determine $\Sigma_\star$, we use the profile output from the MESA model generated at the end of step \ref{pre-irrad} (i.e., the model prior to irradiation). The profile output provides the variation of pressure, $P_i$, temperature, $T_i$, mass interior, $m_i$, and distance from the center, $r_i$, in each radial zone within the planet envelope (indexed by $i$).
The mass column density, $\Sigma_i$, above the $i$th radial zone in the planet envelope is calculated from,
\begin{equation}
\Sigma_{\star, i} = \frac{m_{1} - m_{i}}{4 \pi r_i^2},
\label{eq:column2}
\end{equation}
\noindent where $m_1$ is the total mass at the top of the atmosphere. Making the approximation that once the irradiation is applied to the model planet the optically thin regions of the irradiated planet's atmosphere will be nearly isothermal  at a temperature of $T_{\rm eq}$, we interpolate within the mean opacity tables of \citet{2014ApJS..214...25F} (with a blackbody weighting temperature of $T_{\rm eff,\star}$, and $\left[M/H\right]=0$) to obtain $\kappa_{v,i}$ in each radial zone (i.e., at a pressure $P_i$ and temperature $T_{\rm eq}$). $\Sigma_{\star}$ is determined by solving for the zone at which Eqn.~\ref{ColumnDepth} is satisfied. If the intersection point is unresolved, we take the column depth at the outermost layer, using Eqn.~\ref{eq:column2}. We find $\Sigma_\star$ ranging from 10.0 to $100.0~\mathrm{cm^2\,g^{-1}}$ with the majority of models around 20.0 $\mathrm{cm^2\,g^{-1}}$. We self-consistently determine an initial value of $\Sigma_\star$, and then keep $\Sigma_\star$ constant throughout the planet's evolution. Future work could update $\Sigma_\star$ in time as the planet evolves.

\subsection{H/He Mass Loss}\label{sec:H/He Mass Loss}
We implement the atmospheric mass loss prescription of \cite{2015ApJ...807....8H} in MESA, using the \verb|use_other_adjust_mdot| hook. For the close-orbiting planets that we consider, mass loss is primarily driven by EUV radiation from the host star. Previous studies of planet mass loss with MESA have solely considered hydrodynamic escape wherein the escaping gas has the same composition as the planet's envelope (so the envelope composition does not vary in time). In contrast, following \cite{2015ApJ...807....8H}, we account for diffusive separation of hydrogen and helium and preferential loss of hydrogen.

The planet's energy-limited mass-loss rate, $\Phi_{\rm EL}$ (with dimensions of mass per unit time), is given by,

\begin{equation}
\Phi_{\rm EL} = \frac{L_{\rm EUV}\eta a^2 R_{h}^3}{4Kd^2GM_{p}}
\label{HeliumEq}
\end{equation}

\noindent \citep[e.g.,][]{
1538-4357-598-2-L121,
0004-637X-621-2-1049,
2007A&A...472..329E,
2009ApJ...693...23M}. In Eqn.~\ref{HeliumEq}, $L_{\rm EUV}$ is EUV luminosity of the host star, $\eta$ is the heating efficiency factor (i.e., the fraction of the EUV energy absorbed that goes into unbinding the outer layers of the planet envelope). The factor $a$ is the ratio between radius where EUV photons are absorbed and the planet's homopause radius. We define the homopause radius as $R_h$, and the planet transit radius as $R_p$. We elaborate the calculation and application of $R_h$ and $R_p$ in \S~\ref{sec:radus}. Throughout this work we assume constant values for $\eta$ and $a$ of 0.10 and 1.0 respectively, following \cite{2015ApJ...807....8H}. Finally, $K$ is a correction for the tidal effect of the planet's Roche Lobe.

The Roche potential reduction factor is calculated following \cite{2007A&A...472..329E},
\begin{equation}
K(\epsilon) =  1 - \frac{3}{2\epsilon} + \frac{1}{2\epsilon^{3}},
\label{RocheLobe}
\end{equation}

\noindent where,

\begin{equation}
\epsilon = \left( \frac{M_{p}}{3 M_{\star}} \right)^{{1}/{3}} \frac{d}{R_{h}}.
\label{Epsilon}
\end{equation}

To model the EUV luminosity of the host star, we adopt the empirical relation of \cite{2011A&A...532A...6S}, who studied 80 stars with spectral types from M to F. They found that EUV luminosity is inversely proportional to star age $\tau$,

\begin{equation}
\log_{10}(L_{\rm EUV}) = 22.12 - 1.24\log_{10}(\tau)
\label{Luminosity}
\end{equation}

\noindent In Eqn.~\ref{Luminosity}, $\tau$ is expressed in units of Gyrs and $L_{\rm EUV}$ has units of $\mathrm{J\, \ s^{-1}}$.

The energy-limited escape rate (Eqn.~\ref{HeliumEq}) overestimates mass loss, as thermal and translational energy is carried away by escaping gas \citep{Johnson_2013}. In the transonic regime, the reduction factor ($f_{r}$) is proportional to the ratio of the net EUV heating rate ($Q_{net}$) to the critical heating rate ($Q_{c}$). We adopt $Q_{net}$ and $Q_{c}$ from \cite{2015ApJ...807....8H}, 

\begin{equation}
Q_{net} = \frac{\eta L_{EUV} R^2_h}{4 d^{2}} 
\label{eq:Qnet}
\end{equation}

\begin{equation}
Q_{c} = \frac{4 \pi R_{h} \gamma U(R_{h})}{c_{c} \sigma_{c} K n_{m}} \sqrt{\frac{2 U(R_{h})}{\mu}}
\label{eq:Qc}
\end{equation}

\noindent The collisional mean free path of a particle divided by the scale height (the Knudsen number, $Kn_m$), the heat capacity ratio of the atmosphere ($\gamma$), and the collisional cross section ($c_{c}\sigma_{c}$) are set to 1, 5/3, and $5\times10^{-20}\mathrm m^{2}$ respectively \citep{Johnson_2013}. Our assumption that $Kn_m\approx1$ holds as long as heat is absorbed throughout the atmospheric profile, interior to $R_{h}$. The reduction factor, $f_r$, for transonic and subsonic escape flow is defined as,\\

If $Q_{net} > Q_{c}$,
\begin{equation}
f_{\rm r} \sim \frac{Q_{c}}{Q_{net}}
\end{equation}

Else,
\begin{equation}
f_{r} = 1
\end{equation}

\noindent We adjust the escape rate, $\Phi$, to account for the reduction in mass loss relative to the energy limited assumption.
\begin{equation}
\Phi = f_{r} \Phi_{EL},
\label{eq:f_r}
\end{equation}

The mass escaping from the planet may be explicitly separated into its elemental constituents,
\begin{equation}
    \Phi = \Phi_{\rm H} + \Phi_{\rm He} = 4\pi R_h^2\left(\phi_{\rm H}m_{\rm H}+\phi_{\rm He}m_{\rm He}\right),
    \label{eq:Phiphi}
\end{equation}

\noindent where  $\phi_{\rm H}$ and $\phi_{\rm He}$ represent the number fluxes (in particles per unit area per unit time), and $m_{\rm H}$ and $m_{\rm He}$ are the atomic masses of hydrogen and helium.
Hydrogen is expected to be primarily in atomic (as opposed to molecular) form as it escapes. Eqn.~\ref{eq:Phiphi} neglects escape of any elements heavier than helium. We return to this approximation in \S~\ref{sec:disc}. 

The diffusion limited particle flux, $\phi_{\rm DL}$, mediated by the momentum exchange between hydrogen and helium, from \cite{2015ApJ...807....8H} is,

\begin{equation}
\phi_{\rm DL} = \frac{G M_p (m_{\rm He} - m_{\rm H}) b'}{R_h^2 k T_{H}}
\end{equation}

\noindent where $k$ is the Boltzmann constant, and $b'$ is the effective binary diffusion coefficient (accounting for the partial ionization of hydrogen), and $T_{H}$ is the temperature of the homopause. Following \cite{2015ApJ...807....8H}, we use $T_{H}=10^4~K$ as a conservative estimate of hydrogen-helium fractionation, resulting in a value of b' = 8.0 $\times 10^{20} \ \rm{cm^{-1} s^{-1}}$. The diffusion limited escape rate determines the relative proportions of H and He that escape, 

\begin{equation}
    \frac{\phi_{\rm He}}{X_{\rm He}}=\frac{\phi_{\rm H}}{X_{\rm H}}-\phi_{\rm DL},
    \label{eq:fractionation}
\end{equation}
\noindent where $X_{\rm H}$ and $X_{\rm He}$ represent the mixing ratios (number fractions) of H and He in the atmosphere. 

Solving Eqns.~\ref{eq:Phiphi} and \ref{eq:fractionation}, \cite{2015ApJ...807....8H} derived the following expressions for the hydrogen and helium mass loss rates.

If $\Phi\leq \phi_{\rm DL} X_{\rm H} m_{\rm H} 4\pi R_{h}^2$, 
\begin{eqnarray}
\Phi_{\rm H} &=&\Phi \label{eq:ml1}\\
\Phi_{\rm He} &=& 0
\end{eqnarray}

If $\Phi> \phi_{\rm DL} X_{\rm H} m_{\rm H} 4\pi R_{h}^2$, 
\begin{eqnarray}
\Phi_{\rm H} &=& \frac{\Phi m_{\rm H} X_{\rm H} + \phi_{\rm DL} m_{\rm H} m_{\rm He} X_{\rm H} X_{\rm He} 4\pi R^2_{h}}{m_{\rm H} X_{\rm H} + m_{\rm He} X_{\rm He}}\\
\Phi_{\rm He} &=& \frac{\Phi m_{\rm He} X_{\rm He} - \phi_{\rm DL} m_{\rm H} m_{\rm He} X_{\rm H} X_{\rm He} 4\pi R^2_{h}}{m_{\rm H} X_{\rm H} + m_{\rm He} X_{\rm He}}\label{eq:ml4}
\label{eq:H_He_flux}
\end{eqnarray}

At each MESA time step (indexed by $n$), we store the envelope mass and abundance fractions. We adjust the planet mass through \verb|other_adjust_mdot| in a custom \verb|run_star_extras| file. However, \verb|other_adjust_mdot| may be called multiple times before MESA finds an acceptable time step. In order to avoid multiple changes to the envelope abundances, we set abundance fractions in \verb|extras_finish_step| (in the same \verb|run_star_extras|). This routine is only called at the end of each MESA step.

At the begining of the nth step, H/He mass loss is calculated (according to Eqns.~\ref{eq:ml1}-\ref{eq:ml4}), using a custom \verb|use_other_adjust_mdot|. MESA then attempts to solve the model with these conditions. If MESA accepts the new model, the program proceeds to \verb|extras_finish_step|. Here, we adjust the atmospheric composition with the \verb|extras_finish_step| routine, setting atmospheric abundances through MESA's composition variables \verb|xa(j,k)| for each species and zone.

All MESA envelope models are composed of eight elemental species: \ce{^1H}, \ce{^3He}, \ce{^4He}, \ce{^12C}, \ce{^14N}, \ce{^16O}, \ce{^20Ne}, \ce{^24Mg}. We adjust the proportion of each species at each zone, in response to the hydrogen and helium lost by the planet at each time step $dt$. 

\begin{equation}
X_{n} = \frac{M_{env,n-1}X_{i-1} - (\Phi_{\rm H} * dt)}{M_{env,n-1} - (\Phi_{\rm He} + \Phi_{\rm H}) dt}
\label{frac_X}
\end{equation}

\begin{equation}
Y_{n} = \frac{M_{env,n-1}Y_{n-1} - (\Phi_{\rm He} * dt)}{M_{env,n-1} - (\Phi_{\rm He} + \Phi_{\rm H}) dt}
\label{frac_Y}
\end{equation}

\begin{equation}
Z_{j,n} = \frac{M_{env,n-1}Z_{j,n-1}}{M_{env,n-1} - (\Phi_{\rm He} + \Phi_{\rm H}) dt}
\label{frac_Z}
\end{equation}

\noindent where $Z_j$ is the mass fraction of the $j$th heavy element (i.e., heavier than \ce{^4He}), $M_{env}=M_{p}-M_{core}$ is the total envelope mass, and the subscripts $n-1$ and $n$ refer to initial and final abundance fractions and envelope mass values for each step. The envelope mass is stored throughout a model step, as new abundance fractions are calculated as part of \verb|extras_finish_step|.

\subsection{Planet Radius}\label{sec:radus}
We define planetary radius to be at 1.0 mbar. This is approximately the depth at which the atmosphere becomes optically thin, and is useful as a benchmark transit radius \citep{2009ApJ...702.1413M}. To calculate this, we extrapolate radially from the outermost zone in MESA, assuming a constant scale height. Generally, this increases planetary radius by approximately 10\% - 20\%. However, low mass planets that are highly irradiated can have even larger differences between their transit radii and MESA's outermost zone. The only exception in using 1.0 mbar as a radius definition is when we calculate mass loss rates, explained below.

Above the homopause, the H-He binary diffusion coefficient is greater than the eddy diffusion coefficient ($K_{zz}$) (\citealt{Hu_2012, 2015ApJ...807....8H}). For Eqns.~\ref{HeliumEq}-\ref{frac_Z} we defined the planetary radius as the homopause -- the level in the atmospheres below which the constituent molecules are well mixed.

The binary mixing (molecular diffusion) coefficient between hydrogen and helium is \citep{bird2007transport, doi:10.1021/ie50677a007},

\begin{equation}
    D_{\rm H,He} = \frac{10^{-3} T_{H}^{1.75}\left( \frac{1}{m_{\rm H}} + \frac{1}{m_{\rm He}}\right)^{1/2}}{
    P_{h}\left(\left(\sum_{a} V_H \right)^{1/3} +  \left(\sum_{b} V_{He}\right) ^{1/3}\right) ^{2}}
    \label{eq:DHHe}
\end{equation}

\noindent where $P_{h}$ is pressure at the homopause radius, and $V_{H}$ and $V_{He}$ are the atomic diffusion volumes (\citealt{acp-14-9233-2014, doi:10.1021/ie50677a007, doi:10.1021/ed058pA246.4}).

Like the transit radius, the homopause is not resolved by the top-most zone of our MESA planet models. To calculate the planet radius at the homopause, we thus extrapolate the atmospheric pressure profiles to lower pressures, assuming an isothermal temperature profile (appropriate to the outer radiative zones of these strongly irradiated planets). We find the scale height of the atmosphere using Eqn.~\ref{eq:scale_height}. Equating $D_{\rm H,He}$ and $K_{zz}$ determines the pressure level of the homopause, $P_{\rm h}$, from which the radius at the homopause is estimated using Eqn.~\ref{scale_height},

\begin{equation}
H = \frac{k T_{eq}}{\mu g}
\label{eq:scale_height}
\end{equation}

\begin{equation}
P_h = P_{1}e^{\frac{-Z}{H}}
\label{scale_height}
\end{equation}

\noindent where $g$ is the surface gravity, $\mu$ is the mean molecular mass of the atmosphere, $P_{1}$ is the pressure at the outermost profile zone, and $Z$ is the radius above the outermost zone. We use the value of the scale height at MESA's outermost zone, and extrapolate using constant values for $g$ and $\mu$.

We find that the homopause typically adds between 5\% and 25\% to the planet radius, decreasing as the planet cools and contracts. In the more extreme cases, the homopause could be approximately 50\% larger than the planetary transit radius. The large variability comes from the most extreme of our models, subject to the most intense irradiation with the smallest surface gravity. For planets with masses above 15.0 $M_\oplus$, the homopause is typically 20\% larger than the transit radius. Below the homopause, the atmosphere is well mixed, and the species' mass fractions are constant with depth.

The surface boundary conditions for our MESA models are set through \verb|atm_option| and \verb|atm_T_tau_relation| \citep{2015ApJS..220...15P}. We take the default control option of a simple grey Eddington boundary condition. These specify a set of temperature and pressure conditions for the outermost zone in MESA. However, different surface boundary conditions may be useful in subsequent work.

\begin{figure*}
\centering
    \includegraphics[width=1.0\linewidth]{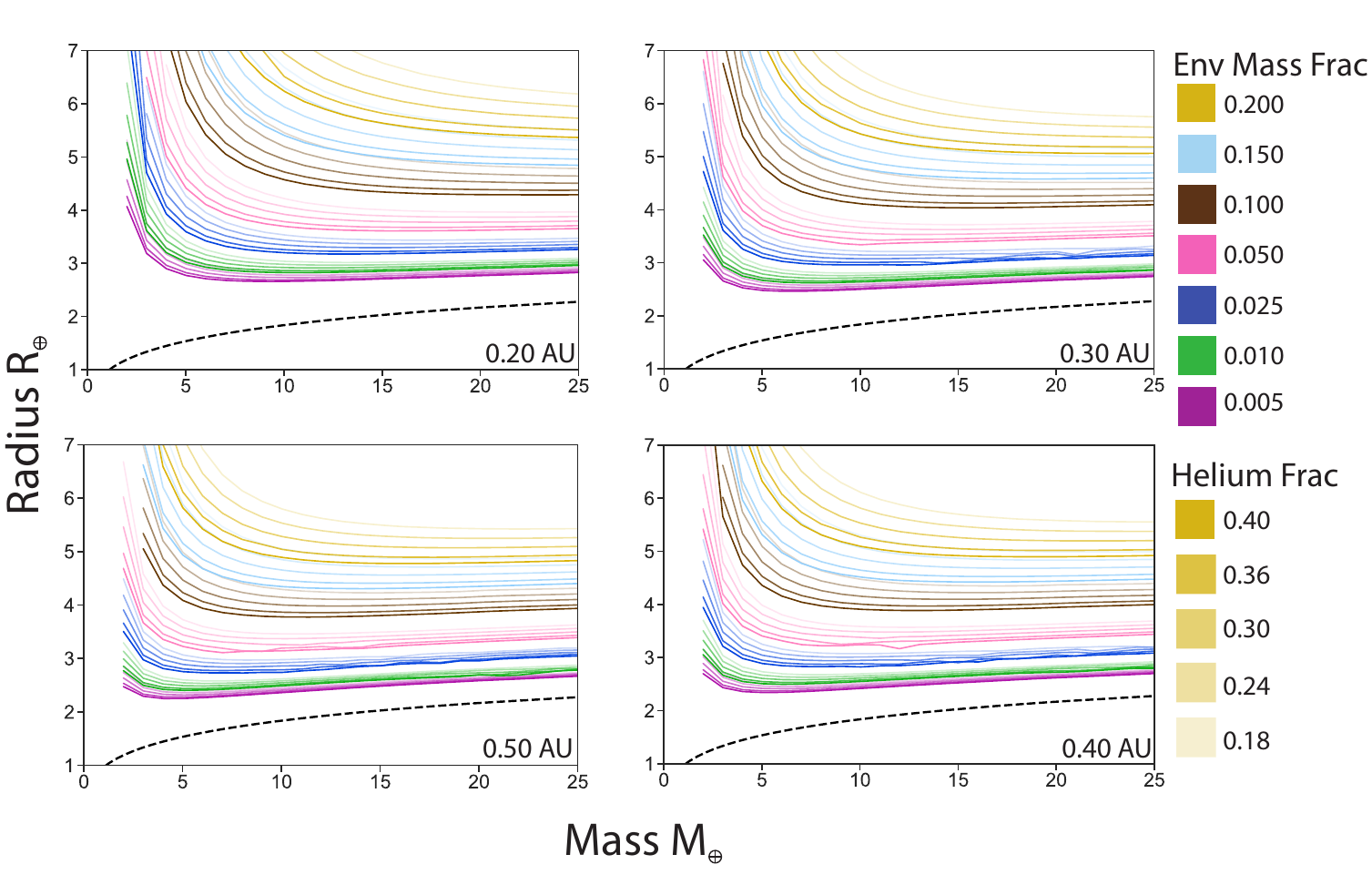}
    \caption{$M_{p}$---$R_{p}$ relations for planets of varying envelope fraction and helium content -- evolved for 5.0 Gyr without mass loss. In each panel we included simulations of a large number of planets with atmospheric helium fractions of 18\%, 24\%, 30\%, 36\%, and 40\% as well as envelope mass fractions of 0.001 (purple), 0.010 (green), 0.025 (blue), 0.05 (pink), 0.10 (brown), 0.15 (light blue), and 0.20 (gold). The darker lines within each color grouping represent higher fractions of initial helium. The dotted black line in all figures represents the planetary core radius. All models were run with a host star temperature of 6000K, and envelope metal mass fraction $Z=0.02$. Differences in helium fraction caused significant radii differences for planets with large ($f_{\rm env}$ > 0.025) envelope fractions. Planets with smaller envelope fractions had much tighter $M_{p}$---$R_{p}$ relations, as envelope added a smaller fraction to the planet's overall radius.}
    \label{Mass_Rad_NoLoss}
\end{figure*}

\subsection{Default Parameters}
Unless otherwise specified, we consider planets orbiting sun-twin host stars (with $M_{\star}=1~M_{\sun}$, $R_{\star}=1~R_{\sun}$, and $T_{\rm eff,\star}=6000$~K). In our GJ~436b case study, however, we adjust the host star properties to those of GJ~436 (see \S~\ref{sec:GJ}).

In all the simulations presented herein, we specified an initial entropy of,

\begin{equation}
S = 7.0 + \frac{M_p}{25.0 \ M_{\oplus}} \ \frac{\rm k_B}{\rm baryon}
\end{equation}

\noindent and allow the planet to evolve for $6\times10^6$~years before applying atmospheric mass loss. Unless otherwise specified, we initialize our planets with solar composition envelopes ($X=0.74$, $Y=0.24$, $Z=0.02$). In determining $\kappa_v$ and $\Sigma_*$, we take the opacity table from \citet{2014ApJS..214...25F} for $\left[M/H\right]=0$. We adopt a Bond albedo of $A=0.2$ to relate $F_\star$ and orbital separation $d$, and set $K_{zz}$ to $10^{9}~\rm cm^{2}~s^{-1}$ throughout.

\section{Results}\label{sec:results}
\subsection{Helium Fraction Dependent Mass-Radius Relations}\label{sec:struc}
We present the first planet mass-radius ($M_p$---$R_p$) relationships that quantify the effect of envelope helium mass fraction, $Y$, on planetary radii (Fig.~\ref{Mass_Rad_NoLoss}). To create the $M_{p}$---$R_{p}$ relations, we evolved models for 10~Gyr over a range of orbital separations (from $d=0.1$ to 0.5~AU), envelope mass fractions ($f_{\rm env}=M_{\rm env}/M_p=0.005$ to 0.20), and helium fractions ($Y=0.18$ to 0.40).

Variations in atmospheric helium mass fraction have a significant effect on planet radius. For planets exterior to 0.2~AU, the effect of $Y$ on planet radii is second only to the effect of $f_{\rm env}$ (within the parameter range explored). An increase in $Y$ from 0.24 to 0.40 corresponds to a decrease in $R_p$ of 0.5\% to $\sim$ 15.0\% (Fig.~\ref{Mass_Rad_NoLoss}). For two similar models (at $M_p=10.0~M_{\oplus}$, $f_{\rm env}=0.05$, and $d=0.20$~AU), we find that a planet with $Y=0.24$ had a 9.2\% larger radius compared to a planet with $Y=0.40$ (3.69 $R_{\oplus}$ compared to 4.06 $R_{\oplus}$) after 10.0 Gyr. The stronger the irradiation and the larger the $f_{\rm env}$, the greater the effect of $Y$ on planet radius, because the envelope contributes a larger overall fraction to the total planet radius.

$Y$ influences the planet radii in Fig.~\ref{Mass_Rad_NoLoss} at a level that may be detected by transit surveys. In the era of {\it Gaia}, planet transit radii may be commonly determined to $\sim$3\% \citep{2017AJ....153..136S}. Across a wide range of $d$, $f_{\rm env}$, and $M_p$, planets with an atmospheric helium fraction of $Y=0.40$ are between 0.5\% and 7.5\% smaller than similar planets with a solar composition ($Y=0.24$). 
Differences in helium content should not be discounted when considering radius measurements, and present an additional dimension of planet compositional diversity that to date has largely been neglected.

\begin{figure*}
\centering
    \includegraphics[width=1.0\linewidth]{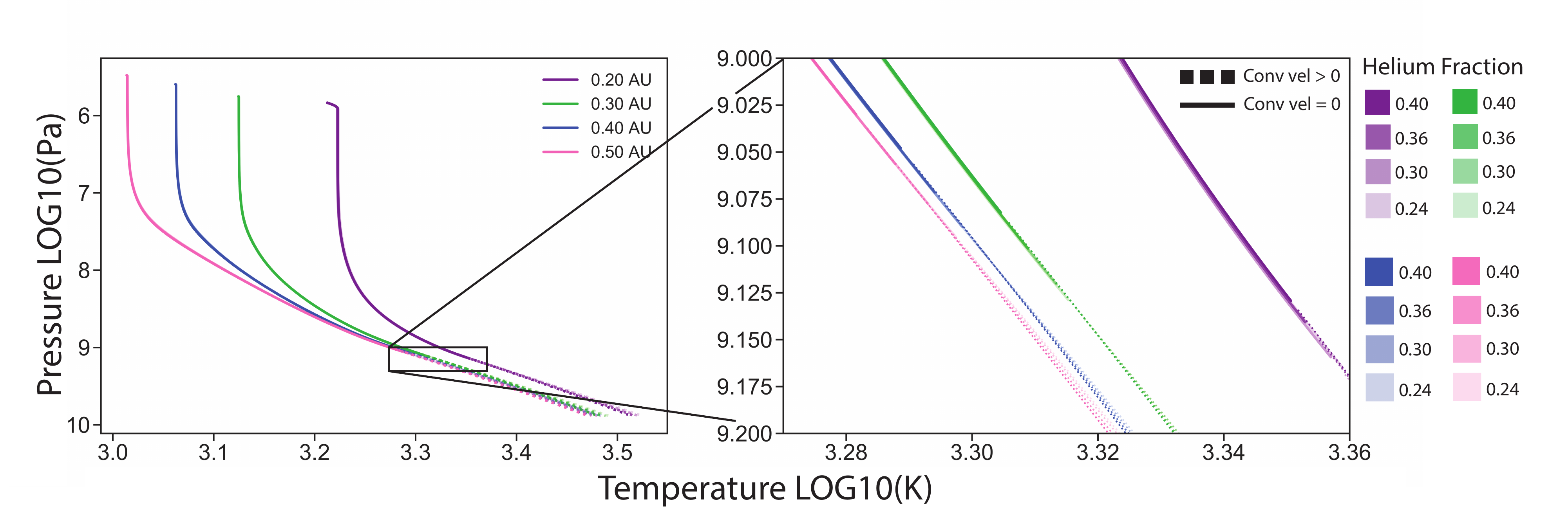}
    \caption{Pressure - Temperature profiles for sixteen planetary models, at varying orbital separations and initial helium fraction. The models were run without mass loss, and were all evolved for 500 Myr. Each model was run at 10.0 $M_{\oplus}$, envelope fraction = 0.01, and $\eta$ = 0.10. The four orbital separations are, from left to right, 0.50, 0.40, 0.30, and 0.20 AU. These correspond to the pink, blue, green and purple lines respectively. Planet envelopes with larger $Y$ have smaller convecting zones. Planets with helium fractions above $Y=0.24$ had envelopes which were entirely radiative by the age of $\sim$ 5.0~Gyr.}
    \label{Temp-Pressure}
\end{figure*}

We note that atmospheric escape will not increase the ratio of hydrogen relative to helium in primordial planetary envelopes. However, other atmospheric sources such as outgassing and cometary delivery of volatiles will preferentially contribute hydrogen and not helium \citep[e.g.,][]{ElkinsTanton&Seager2008bApJ, Rogers&Seager2010bApJ, RogersEt2011ApJ}. Though not the primary emphasis of this paper, we include models with $Y=0.18$ in Fig.~\ref{Mass_Rad_NoLoss} to highlight how, in theory, a decrease in the helium mass fraction below solar proportions would affect the $M_p-R_p$ relations. In section~\ref{sec:loss} we investigate what range of planet configurations (mass, envelope composition, envelope mass fraction) can be achieved by escape from initially solar composition primordial envelopes.

\subsection{Effect of Helium on Envelope P--T Structure}
Physically, the helium mass fraction has multiple effects on the interior structure of a planet's envelope that in turn affect the planet's transit radius. Figure~\ref{Temp-Pressure} presents atmospheric pressure-temperature ($P-T$) profiles for planets having various values of $Y$.

The larger the concentration of helium in a planet's envelope, the higher the mean molecular weight (for a given envelope metallicity, $Z$). As a result, helium enhanced planets have smaller atmospheric scale heights (Eqn.~\ref{eq:scale_height}), and smaller radii (as observed in Fig.~\ref{Mass_Rad_NoLoss}). 

Changing the relative mass fractions of hydrogen and helium also affects the adiabatic temperature gradient in the convection zones of planetary envelopes, because hydrogen has a larger heat capacity than helium. Based on the \cite{1995ApJS...99..713S} equations of state, the specific heat capacity (per unit mass), $c_P$, of hydrogen is 2.7 -- 5.6 times larger than that of helium at pressures of $10^{3.0}$ to $10^{10}$ Pa and temperatures of $10^{2.5}$ to $10^{3.3}$ K.  Molecular dissociation of hydrogen begins at $\sim 2500$ K, further increasing the specific heat capacity of hydrogen to as much as 50 times that of helium. However, the majority of our atmospheric profiles do not encounter temperatures this high. The adiabatic temperature gradient depends on the heat capacity as in Eqn.~\ref{eq:adiabatic},

\begin{equation}
\label{eq:adiabatic}
\left(\frac{dT}{dP}\right)_S = \frac{\alpha_v T}{\rho c_P},
\end{equation}

\noindent where $\rho$ is the density, and $\alpha_v=1/v (\partial v/ \partial T)_p $ is the coefficient of thermal expansion. The lower heat capacities of hydrogen-depleted/helium-enhanced envelopes in turn lead to steeper temperature gradients in their convecting regions (i.e., with larger changes in temperature with pressure) compared to envelopes with solar hydrogen-to-helium ratios. For example, taking two $M_p=10.0~M_{\oplus}$, $d=0.50$~AU, $f_{\rm env}=0.01$ planet models, the $Y=0.24$ and $Y=0.40$ models have a convective lapse rate of $0.24~\rm{K\, km^{-1}}$ and $0.27~\rm{K\, km^{-1}}$, respectively, at a pressure of $10^{9}$~Pa.

In addition to changing the temperature profile in zones where convection is occurring, the $Y$ dependence of the adiabatic lapse rate also influences the location of the radiative-convective boundary. With a steeper adiabatic temperature gradient, helium-enhanced envelopes are more stable against convection and have narrower convecting zones (Fig.~\ref{Temp-Pressure}).

\begin{figure*}
\centering
    \includegraphics[width=1.0\linewidth]{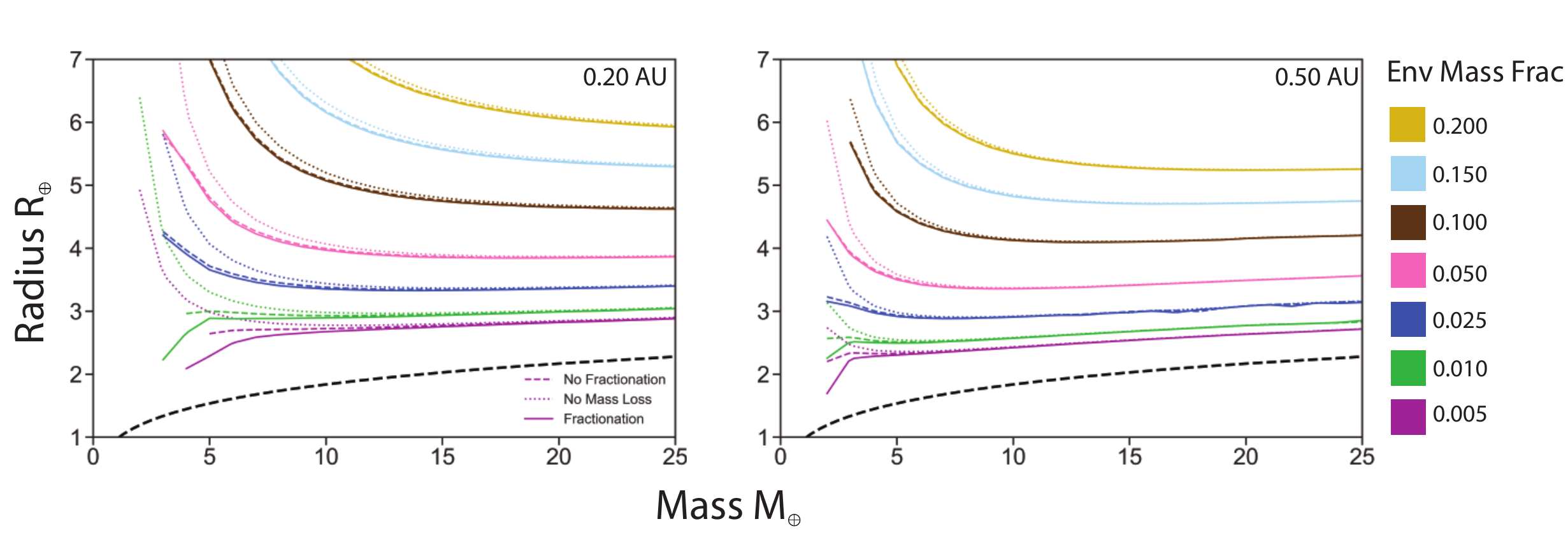}
    \caption{$M_{p}$---$R_{p}$ relations of planetary models with evolved with no mass loss, mass loss without fractionation of atmospheric species, and mass loss with fractionation. Models range from 2.0 $M_{\oplus}$ to 25.0 $M_{\oplus}$, initial envelope fractions from 0.001 to 0.20, and initial abundances of $Y=0.24$, $Z=0.02$, $X=0.74$. Additionally, all models were evolved for 5.0 Gyr around a host star with a temperature of 6000K.
    Though envelope mass fraction is the input factor with the strongest influence on the $M_{p}$---$R_{p}$ relations, the mass loss mode assumed leads to significant radius differences -- particularly for small and highly irradiated planets. The most irradiated planets had envelopes that became unbound at the smallest masses modeled. Below 15.0 $M_{\oplus}$, envelope erosion and helium enhancement dramatically reduced the radii of modeled planets.}
    \label{Mass_Rad_loss}
\end{figure*}

In general, increasing the proportion of helium relative to hydrogen in a planet's envelope lowers the opacity. In the planet models presented herein we have neglected the effect of non-solar $Y$ on opacity; the opacities from \citet{2014ApJS..214...25F} and the \verb|lowT_Freedman11| table implemented in MESA \citep{2008ApJS..174..504F} are only available for solar ratios of hydrogen and helium. We quantify the effect of this approximation in our models in section~\ref{sec:caveats}. We describe below how the decrease in opacity with $Y$ is expected to influence both the optically thin atmosphere and optically thick radiative zones of true planets.

In the optically thin atmosphere, the helium mass fraction of an envelope affects the depth at which the stellar irradiation penetrates into the planetary atmosphere. The higher the helium content in a planet envelope, the lower the wavelength-averaged (irradiation mean) opacity to incoming starlight, $\kappa_v$. As a result, the irradiation from the host star penetrates further into helium-enhanced planetary atmospheres.

Within optically thick radiative zones, helium enhanced planets with lower Rosseland mean opacities can more easily transport energy by radiative diffusion and will have shallower radiative temperature gradients. Based on the \verb|lowT_fa05_gs98| opacity tables \citep{FergusonEt2005ApJ}, increasing the $Y$ from 0.18 to 0.36 decreased the Rosseland mean opacity by 5--15\%, depending on temperature and density. In addition to changing the temperature profile in optically thick radiative zones, the $Y$ dependence of opacity would lead helium-enhanced envelopes to be more stable against convection and to have narrower convecting zones.

\subsection{The Effect of Mass Loss on Atmospheric Structure and Composition}\label{sec:loss}

In the previous two sections, we explored how varying the $X$ and $Y$ of a mini-Neptune could affect the planet's radius and atmospheric structure. Now we turn to considering a mechanism by which planets may become enhanced in helium relative to hydrogen.

To investigate diffusion modulated atmospheric escape as a pathway for creating helium enhanced planets, we simulated a grid of planets with varying masses, orbital separations, and envelope mass fractions for three different mass loss regimes. Figure \ref{Mass_Rad_loss} shows the resulting $M_p$ --- $R_p$ relations after 5.0 Gyr of evolution. Models ranged from 2.0 $M_\oplus$ to 25.0 $M_\oplus$, 0.001 to 0.20 $f_{\rm env}$, initial abundances of Y=0.24, Z=0.02, X=0.74, and were evolved around a 6000K host star.

In Figure \ref{Mass_Rad_loss} we show three simulated regimes of mass loss. First, we simulated the mass loss regime from \cite{2015ApJ...807....8H}. Fractionation between hydrogen and helium leads to preferential hydrogen loss. Next, we computed evolution tracks for planets with mass loss but not fractionation. The rate of mass loss was calculated in the same manner as in \cite{2015ApJ...807....8H}. These planets had constant envelope compositions throughout their evolution. Last, we modeled a third set of planets without mass loss.

We find that preferential hydrogen loss can shape the $M_p$--$R_p$ relations of sub-Neptune-mass planet populations. Planets simulated with fractionation often had atmospheric helium mass fractions significantly higher than their initial abundance of $Y=0.24$. We found $Y$ fractions as high as 0.35 within our simulated grid, and would expect to find even more helium enhancement if we simulated denser grid of highly irradiated planets with small initial envelopes.

Low mass $\left(M_p\lesssim10~M_{\oplus}\right)$, highly irradiated planets with envelope fractions below 1.0\% showed the largest differences in radii between planets evolved with fractionation mass loss and planets evolved with non-fractionation (constant envelope composition) mass loss. These effects can be seen in Figure \ref{Mass_Rad_loss}. In this low-mass, low-$f_{\rm env}$ regime, planets  run with mass loss are significantly smaller than planets run without mass loss. Planets that evolve with preferential hydrogen loss are smaller still than planets run with constant composition mass loss. Over the rest of the parameter space explored, envelope mass fraction dominates the $M_p$--$R_p$ relations, being the most important input factor affecting planet radii at 5.0 Gyr.

\subsection{GJ 436b}\label{sec:GJ}
\citet{2015ApJ...807....8H} first proposed the possibility of helium-enhanced mini-Neptunes to explain the lack of CH$_4$ observed in the emission spectrum of GJ 436b. GJ 436b is a transiting Neptune-size planet ($4.22\substack{+0.09 \\ -0.10}~R_\oplus, ~ 23.17\pm~0.79~M_\oplus$; \citet{2007ApJ...671L..65T}) with an orbital semi major axis of 0.02872~$\pm$~0.00027~AU. Its host star is type M2.5V ($M_{\star}=~0.452\substack{+0.014 \\ -0.012}~M_\odot$, $R_{\star}=~0.464\substack{+0.009 \\ -0.011}~R_\odot$; \citet{2007ApJ...671L..65T}), and has an effective temperature of $T_{\rm eff,\star}=3350\pm~300$~K \citep{2007ApJ...667L.199D}). There is significant uncertainty on GJ 436b's age. \cite{2018Natur.553..477B} places the age of the system between 4-8 Gyr. For the purposes of our model, we assume GJ 436b has an age of 5.0 Gyr unless otherwise specified. This is well within the age range where helium enhancement is possible. \citet{2015ApJ...807....8H} theorized that if hydrodynamic mass loss was to significantly alter the atmospheric composition of GJ 436b, the planet must begin evolution with the presence of an initial H/He envelope fraction less than or equal to $10^{-3}$ of the planetary mass.

\begin{figure*}
\centering
    \includegraphics[width=1.0\linewidth]{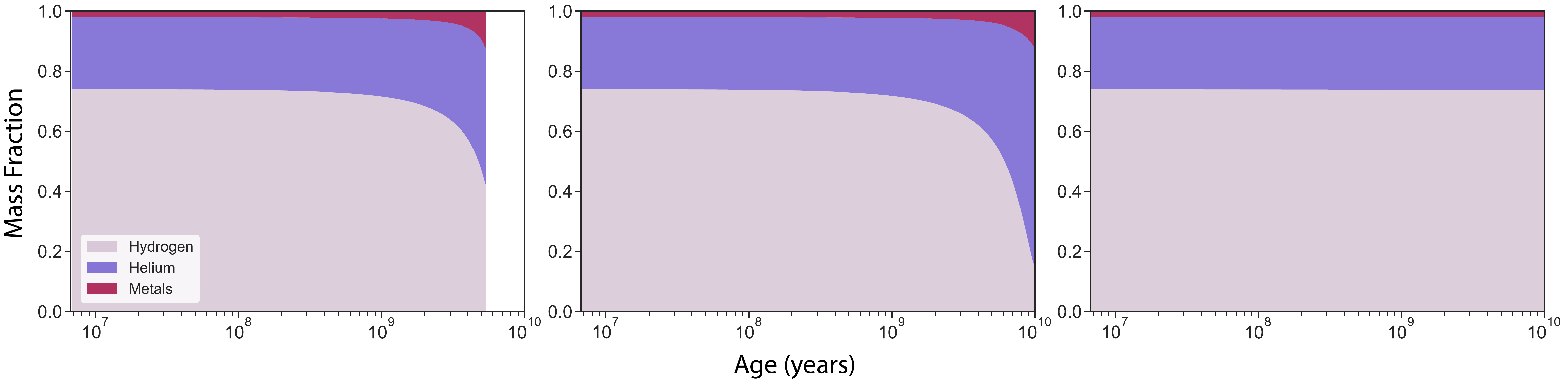}
    \caption{Plot of changing atmospheric abundances throughout the 10.0 Gyr evolution of three planetary models. All three models were run with hydrodynamic mass loss, and began with $M_{p} = \rm{23.5}~M_{\oplus}$, and initial envelope fractions of 0.003. From left to right, the planets were run with orbital separations of 0.01, 0.03 and 1.00 AU. From left to right we can see the different extremes of mass loss. First, a planet that was so irradiated it failed the full 10.0 Gyr evolution. Before failing, this planet had become basically a remnant core, with no envelope remaining. Second, a model that became helium enhanced. A significant fraction of the total atmospheric hydrogen has been lost. Last, no significant composition change occurred throughout the planet evolution at 0.50 AU. The choice of $M_{p} = \rm{23.5}~M_{\oplus}$ for each of the three models shows specifically how the magnitude of irradiation flux affects the mass loss evolution of models similar in mass, radius, and orbital separation radius to GJ 436b.}
    \label{Pathways}
\end{figure*}

We find that GJ 436b cannot be significantly enhanced in helium or depleted in hydrogen compared to its host star via the mechanism proposed by \citet{2015ApJ...807....8H}. The radius and orbital separation of GJ 436b places it outside the realm of possible helium enhancement that we modeled. GJ 436b is too large, and requires too massive a primordial envelope, to become sufficiently helium-enhanced via atmospheric mass loss. We ran a large suite of models with characteristics similar to GJ 436b and its host star (Fig.~\ref{GJ_Ages}), and found that GJ 436b was significantly outside the domain of helium enhancement. This conclusion is robust against small changes in the mass, radius, orbital separation, and age of GJ 436b. In our modeled grid, no planet with a radius above 3.00 $R_{\oplus}$ reached helium fraction above $Y=0.40$ within 10~Gyr. Furthermore, 23.0 - 24.0 $M_\oplus$ models from 0.01 to 1.00 AU and with final radii such that 4.0~$R_{\oplus}$ < $R_{p}$ < 5.0~$R_{\oplus}$, all had final envelope fractions between 0.05 and 0.20 -- an order of magnitude larger than was typical for helium enhancement. Envelope fractions this large preclude hydrodynamic mass loss from significantly changing a planet's atmospheric composition. GJ 436b's radius of 4.22 $R_\oplus$ necessitates the presence of a significant gaseous envelope and places it outside the domain of envelope mass-fraction parameter space for which helium enhancement is possible.

We found good agreement between our simulated mass loss rate for GJ 436b and previous work. For a 23.5 $M_\oplus$ model run at 0.026 AU and with a final radius of 4.21 $R_\oplus$, we found a mass loss rate of $1.98\times10^{9}~\mathrm{g~s^{-1}}$ at 5.0 Gyr. In comparison, \cite{2015ApJ...807....8H} modeled the escape rate of GJ 436b at $10^{8}$~--~$10^{10}~ \mathrm{g~s^{-1}}$. Lyman-$\alpha$ transit transmission spectra show a current mass loss rate of $10^{8}$~--~$10^{9}~ \mathrm{g\, s^{-1}}$ for GJ 436b \citep{2015Natur.522..459E}. Furthermore, \cite{2015Natur.522..459E} show that the escape rate of GJ 436b would have been significantly greater during the earlier evolution of its host star. Overall, our models agree with \cite{2015ApJ...807....8H}, as well as observational estimates of GJ 436b's loss rate.

\subsection{Helium Enhancement in GJ 436b Mass Planets}
\begin{figure*}
\centering
    \includegraphics[width=1.0\linewidth]{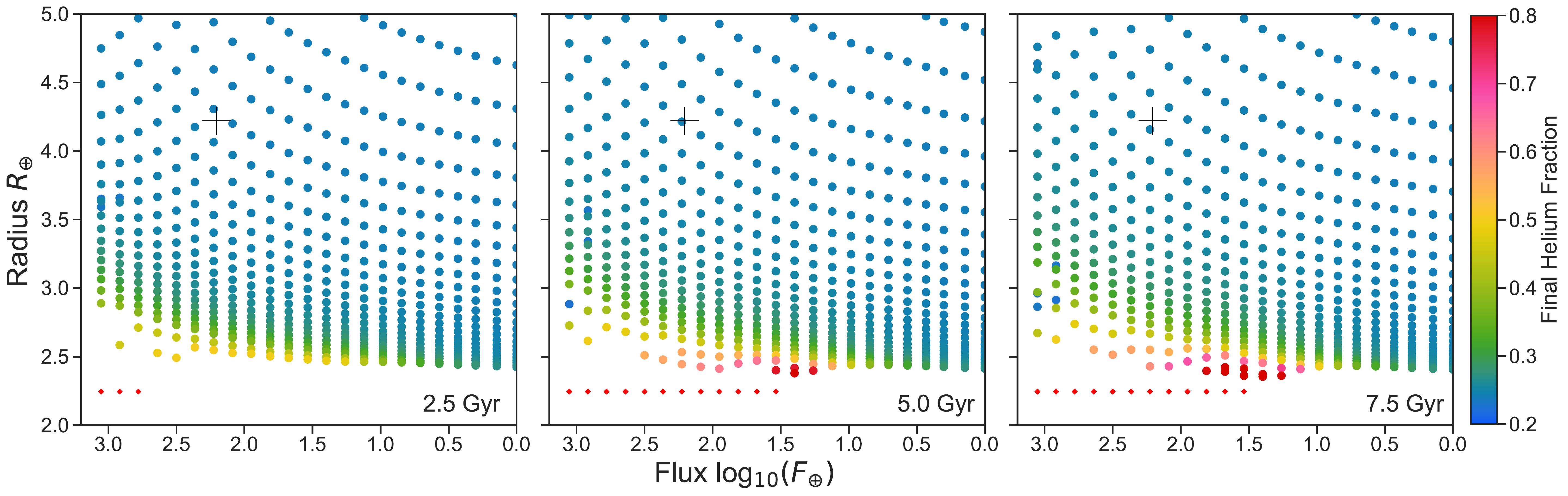}
    \caption{The Flux---Radius distribution of models evolved around a GJ 436-like star, with the presence of hydrodynamic H/He loss. The above models had masses an initial mass 23.5 $M_\oplus$, orbital separations from 0.01 to 1.00 AU, and $f_{\rm env}$ between 0.001 and 0.200. Above are the distributions at 2.5, 5.0, and 7.5 Gyr. The black cross corresponds to the flux and radius of GJ 436b, solidly in a parameter space where helium enhancement did not occur. However, we see significant helium enhancement along the lower radius boundary after 5.0 Gyr of mass loss. The red Xs correspond to remnant core models which failed to evolve for the time frame plotted, and whose envelopes became unbound by the incident flux. We assigned these models radii corresponding to the core radius at 23.5 $M_\oplus$. These models constrain the progression of helium enhancement for a GJ 436b-mass planet, and the parameter space for which final Y fractions might exceed 0.40.}
    \label{GJ_Ages}
\end{figure*}

Although we find that GJ 436b cannot be significantly helium enhanced, it is possible that exoplanets with smaller initial envelope fractions could be. We ran an extensive grid of simulations for planets with similar masses to GJ 436b. For Fig.~\ref{GJ_Ages}, we evolved planets from 0.01 to 1.00 AU, 0.001 to 0.20 $f_{\rm env}$, and masses from 23.0 to 24.0 $M_\oplus$, around a 3350K, star. In future work we will explore helium enhancement for a larger range of sub-Neptune-mass models.

A key takeaway of our simulations of planet evolution is that helium enhancement is possible. Preferential escape of light gases cause planet atmospheres to increase in helium and metals relative to hydrogen over billions of years. Through mass loss, planets can progress from solar helium abundances to having atmospheric helium mass fractions greater than $Y=0.40$. We describe this in detail below.

The outcomes of our evolution simulations can be divided into three categories based on helium enhancement and envelope erosion. Figure~\ref{Pathways} shows the evolution of three planets -- i.e., one from each category. First, for planets with large initial envelopes ($>1.0\%$), or at distances further than 0.50 AU, there was no significant helium enhancement. These planets did not experience significant mass loss, and obeyed the same $M_{p}$---$R_{p}$ relations shown in Figure~\ref{Mass_Rad_NoLoss}. Their final radius is largely determined by their initial envelope mass fraction and mass. Second, we found planets that had progressed in atmospheric helium abundance from $Y=0.24$ to $Y$ in excess of $0.40$ over a timescale of several Gyr. These planets were along the lower radius boundary of the planets that retained their H/He envelopes for a given mass, and were highly irradiated. Last, the most highly irradiated planets became remnant cores through envelope erosion, and had envelopes that became unbound before 10 Gyr had passed.

Models exterior to 1.00 AU or with initial envelope fractions greater than 1.0\% did not become helium enhanced over 10.0 Gyr. The rate of helium mass loss relative to hydrogen was not large enough to significantly change the composition of these planets. The physical structure of these models was similar to previous sub-Neptune interior structure models \citep[e.g.,][]{Chen&Rogers2016ApJ, 2014ApJ...792....1L, 2015ApJ...808..150H, ValenciaEt2007bApJ}.

After 5.0 Gyr of mass loss, we found that helium enhanced models had final radii between 2.34 $R_{\oplus}$ and 2.90 $R_{\oplus}$. These models were among the most highly irradiated, and had some of the smallest initial envelope fractions.  Helium enhanced models lost between 35\% and 90\% of their initial envelope mass. Furthermore, they had irradiation fluxes from 7.0 to 1130.0 $F_{\oplus}$, initial envelope fractions below 0.0047, final envelope mass fractions below 0.0025, Z fractions from 0.033 to 0.225, and transit radii from 5\% to 29\% above that of their rocky cores.

The progression of helium enhancement for 23.5 $M_{\oplus}$ models can be seen in Figure \ref{GJ_Ages}.  Planets initially lose helium and hydrogen in relatively equal proportions, and we do not find planets with Y $\geq$ 0.40 before approximately 2.5 Gyr. As planets continue to evolve helium enhancement becomes a prominent feature along the lower radius boundary. However, after 5.0 Gyr, changes in atmospheric composition and planet radius largely abate. This is due to a decrease in mass loss rates and a decrease in envelope contraction, respectively.

Last, the most highly irradiated planets had envelopes which became completely unbound in our simulations. We define these remnant cores as models that failed to evolve for the full 10 Gyr, and lost more than 75\% of their initial envelope before failing. We manually assigned these planets radii equal to their rocky cores, and placed them in the appropriate location within the flux-radius relations of Figure \ref{GJ_Ages}. They occupy a parameter space within the F-R chart along the lowest radius boundary, and more strongly irradiated than models that evolved the full 10.0 Gyr.

We found remnant cores for planets which were irradiated at rates of at least 30.0 $F_{\oplus}$. The larger the planet mass, the larger the surface gravity, and the more flux necessary to evaporate the envelope. These remnant cores show a clear demarcation between planets that had managed to retain some of their envelope, and planets that had lost their entire envelope due to hydrodynamic mass loss.

\section{Discussion}\label{sec:disc}
\subsection{Helium enhancement in tightly orbiting sub-Neptunes}\label{sec:discussion}

Under irradiation approximately $10^{1}$--$10^{3}$ times that of Earth, helium enhancement was possible for planets with masses similar to GJ 436b. We found helium enhanced planets primarily along the lower radius boundary of all planets that had retained a gaseous H/He envelope (Fig.~\ref{GJ_Ages}). These planets usually had final radii 10-25\% larger than that of their rocky core. Helium enhancement necessitates that planets begin their evolution with relatively small envelope fractions. Compounding this, helium enhanced planets ($Y\geq 0.40$) further lost a minimum of 35\% of their envelope mass through envelope erosion. Planets with large envelope masses do not lose enough mass  to significantly change their atmospheric abundances. All models that eventually became helium enhanced began their evolution with $f_{\rm env} \leq 0.005$. As a result, helium enhancement occurs within a narrow range of simulated planet radii (Fig.~\ref{GJ_Ages}).

Helium enhancement appeared as a prominent feature after approximately 2.5 Gyr of hydrodynamic mass loss. Planets with small envelope fractions ($f_{\rm env} \leq 0.0025$) become helium enhanced in the shortest time frame. After several Gyr however, larger-envelope planets undergoing steady mass loss with diffusive separation of hydrogen and helium can become helium enhanced. Figure~\ref{GJ_Ages} shows a set of 23.5 $M_\oplus$ models at various evolution snapshots to show the progression of helium enhancement. Helium enhancement occurs first along the lower radius boundary, and subsequently spreads to planets with larger envelopes and larger radii.

The \cite{2015ApJ...807....8H} model that we implemented treated the loss of hydrogen and helium but not metals. Therefore, complete envelope erosion was not possible. Additionally, our model assumed constant values for $\eta$ and the EUV absorption radius ratio ($a$) in Eqn.~\ref{HeliumEq}. Future work could expand on our mass loss routine, and include metal loss and better treatment of the most extreme edge cases.

Although GJ 436b is not significantly helium enhanced, atmospheric mass loss may significantly alter the envelope compositions of similar mass planets with smaller initial envelopes. At 5.0 Gyr, GJ 436b mass planets (23.5 $M_\oplus$) with $Y \geq 0.40$ had final radii between 2.38~$R_\oplus$ and 2.84~$R_\oplus$ and $F_{\star} \gtrsim 10~F_{\oplus}$ (Fig.~\ref{GJ_Ages}). The final helium mass fraction gradually declines (approaching the primordial solar value assumed) as planetary radius increases, and irradiation flux decreases. In the population of GJ 436b mass planets we modeled, only models with radii below 3.00 $R_\oplus$ had helium abundances above $Y=0.40$. Furthermore, given sufficient irradiation, we expect to find helium enhancement in the broader sub-Neptune population as well.

\subsection{Metallicity Enhancement}
In addition to engendering super-solar proportions of helium, preferential loss of light gases may lead to metallicity enhancements in sub-Neptune atmospheres. In our simulations, the planets that became enhanced in helium also became enhanced in metals. The planets along the lower (small-radius) boundary of Figure~\ref{GJ_Ages} experienced significant atmospheric metallicity enhancement (relative to their initial solar metallicity starting envelope composition, $Z=0.02$), achieving final metal mass fractions between $Z=0.04$ and 0.23.

Theoretical work by \citet{FortneyEt2013ApJ} has pointed to links between planet atmospheric metallicity and formation process (e.g., the size distribution of planetesimals, and the mass of H/He gas accreted). Our results highlight that evolution (specifically atmospheric escape) in addition to the formation process could impact planet atmospheric metallicities -- contributing (over several Gyr) to Neptune-mass and sub-Neptune-mass planets having more metal--rich atmospheres than their massive jovian cousins. Though beyond the scope of this paper, expanding the model to include the loss of heavy elements could be a future endeavor.

\subsection{Model Caveats}
\label{sec:caveats}
Some of our planet models failed to evolve for the full 10.0 Gyr evolution due to exceeding the $\rho-T$ boundary limits in the EOS module of MESA. In particular, models with masses $\geq$ 15.0 $M_\oplus$, $f_{\rm env} = 0.01 - 0.10$, and outer envelope temperatures between 750-1500~K failed after approximately 8 Gyr. The low density and low temperature SCVH tables within MESA's EOS module (\cite{1995ApJS...99..713S}) roughly cover calculations for which log($\rho$) - 2 * log(T) + 12 $\leq$ 5.0. As our models cooled and contracted, they pushed the limits of what MESA was able to simulate. Future versions of MESA aim to expand the range of temperatures and densities in the EOS tables, allowing densities from $10^{-8}$ to $10^{6}~{\rm g ~ cm^{-3}}$, pressures from $10^{-9}$ to $10^{13}$ GPa, and temperatures from $10^{2}$ to $10^{8}$ K (\citealt{2019ApJ...872...51C}; Josiah Schwab, priv. communication).

Planets with orbits interior to 0.05 AU and with initial envelope fractions less than 0.003 are likely to become remnant cores. MESA experienced time step convergence issues with these models, and failed to evolve them past approximately 3.0$\times$ $10^{9}$ years. Often, they had final radii in excess of 10 $R_{\oplus}$, suggesting that the atmosphere had become gravitationally unbound before model failure. Additionally, most of the mass lost for these simulations was in the energy limited escape regime. This meant that the remnant cores lost both hydrogen and helium, and at rates approximately equal to their abundances. Therefore, before becoming unbound, they did not become appreciably enhanced in helium relative to hydrogen.

\cite{0004-637X-817-2-107} described this phenomenon, and suggested that rapid atmospheric ``boil off'' could deplete the gaseous envelopes of $M_p$ $\leq$ 10.0 $M_\oplus$ planets over a period of $10^{5}$ years. They point to ``boil off'' as a possible cause of the dearth of Kepler planets with orbits interior to 0.50 AU and radii above 2.50 $R_\oplus$.

Numerically, remnant cores are (not unexpectedly) challenging to model in MESA. These simulations are extremely sensitive to small changes in envelope fraction and orbital separation, and were only found for the most unstable conditions that we modeled. We differentiate these models from the super helium enhanced models that have evolved for the full 10.0 Gyr (Fig.~\ref{GJ_Ages}), and manually assign the remnant cores the radii of their rocky core.

We chose the \verb|lowT_Freedman11| opacity tables to model atmospheric opacity, as they had the most accurate data for the atmospheric temperature ranges that our models spanned. However, the \verb|lowT_Freedman11| opacity tables do not include the effects of varying helium concentrations on atmospheric opacity \citep{2014ApJS..214...25F}. Other opacity tables within MESA, such as \verb|lowT_fa05_gs98|, include the effects of varying $Y$, but have more restrictive temperature ranges \citep{FergusonEt2005ApJ}. The tables from \cite{2014ApJS..214...25F} extend from 75-4000~K, while the tables from \cite{FergusonEt2005ApJ} cover 500-30,000~K. Our models have outer envelope temperatures from approximately 250K - 3000K, necessitating the choice of the \verb|lowT_Freedman11| tables.

To benchmark how the choice of opacity table influences our results, we re-ran the suite of planets in Figure~\ref{Temp-Pressure} with \verb|lowT_fa05_gs98|. Using \verb|lowT_fa05_gs98|, we find that planets had slightly (less than 2.0\%) larger final radii than planets run with \verb|lowT_Freedman11|.

Notably, the level of helium enhancement was almost identical between the opacity table choices. Thus, the choice of opacity table does not change our overall qualitative results: that GJ~436b is not helium enhanced, but that similar planets with smaller envelopes may be.

Our choice of $T_{H}= \rm 10^{4} \rm~K$ results in a conservative estimate of the hydrogen-helium fractionation effect.  \cite{2009ApJ...693...23M} show that above the photoionization base, Lyman-$\alpha$ cooling regulates the temperature to at most $\sim 10^{4} \rm~K$. Further increasing incident UV power is balanced out by larger radiative losses. Choosing $T_{H}= \rm 10^{4}$ -- as opposed to a lower estimate like the planet equilibrium temperature -- decreases fractionation in two ways. First, the coupling between neutral hydrogen and helium increases with temperature. Second, a larger fraction of hydrogen is ionized at higher temperatures. From $10^{3}$ to $10^{4}~$K, $~\rm H^{+}$--He is more strongly coupled than neutral H-He. Therefore, increasing the ionized fraction increases the average coupling weight for all atmospheric H-He (\citealt{Mason1970, schunk1980ionospheres, 2015ApJ...807....8H}). Both effects result in more helium being carried with escaping hydrogen. Even with the conservative estimate however, we found extensive helium enhancement.

\begin{figure}
\centering
    \includegraphics[width=1.0\linewidth]{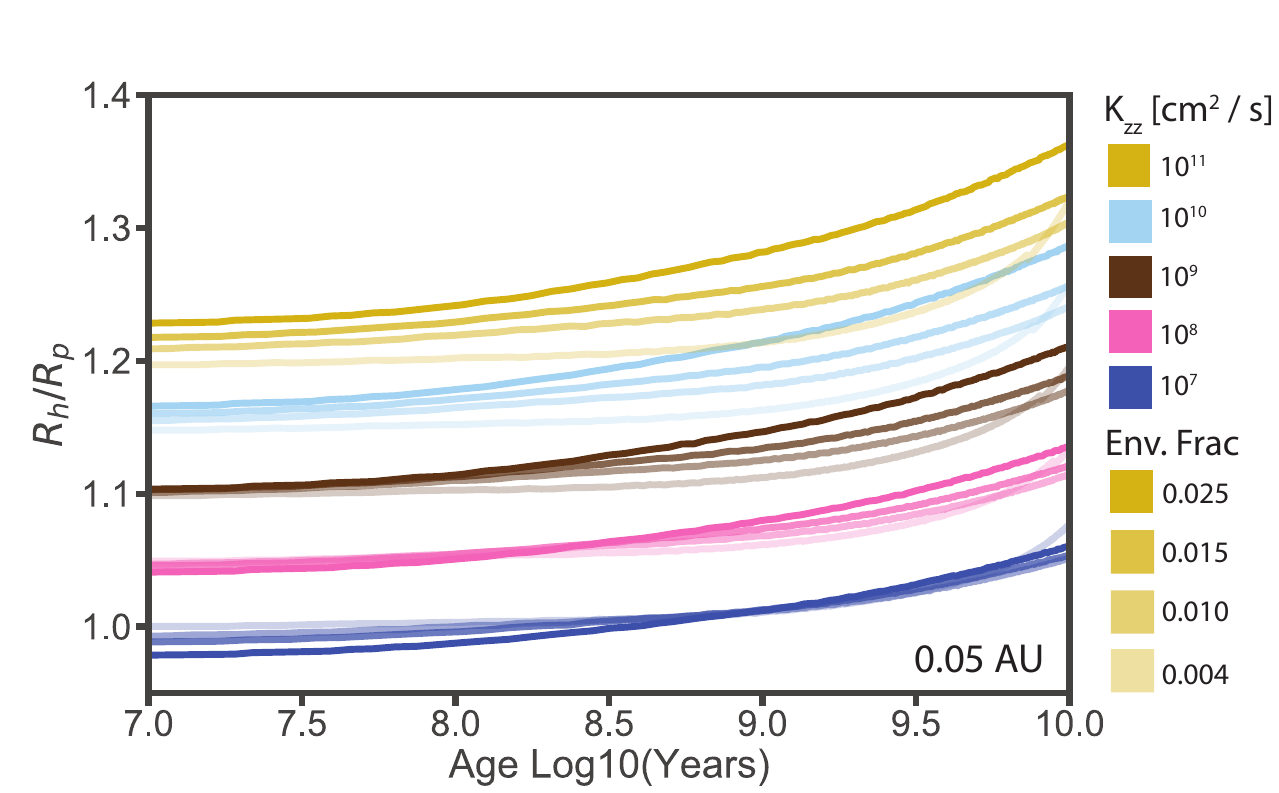}
    \caption{The radius evolution of five sets of planets, with varying values of $K_{zz}$. We simulated the evolution of 25 $M_{\oplus}$ planets for 10.0 Gyr 0.05 AU from its host star, and with initial envelope mass fractions from 0.004 to 0.025. Depending on $K_{zz}$ choice, the homopause is between 10\% below and 40\% above the planet's transit radius.}
    \label{kzz}
\end{figure}

Different parameterizations of $K_{zz}$ can have a significant effect on the homopause radius, as show in Figure \ref{kzz}. We compared five different values of $K_{zz}$, from 1/100th of our chosen value of $10^{9}~\rm cm^{2}~s^{-1}$ to 100 times greater. Smaller eddy diffusion coefficient values resulted in smaller resulting $R_h$ values. However, the increase in radius between the smallest parametarization for $K_{zz}$ and the largest caused less than a 50\% increase in $R_h$.

For planets with masses as large as GJ 436b, variances in the homopause radius will not significantly change how planetary composition evolves. We found significant changes in atmospheric composition across a wide range of orbital separations, and the overall results for helium enhancement present are robust against the relatively small changes that different $K_{zz}$ values would cause for GJ 436b. The homopause radius is only applicable in determining the mass loss rate, and therefore unimportant for planets that never lose a significant portion of their envelope (i.e., $f_{\rm env} > 1.0\%$)).

Future work could better model $\eta$, g, $\mu$, and $K_{zz}$. Differing values of $\eta$ would affect the mass loss rate of all models equally. In contrast, the effects of varying $K_{zz}$ will most signficicantly impact the mass loss rates of small planets with low surface gravity.

\subsection{Observational Consequences of He enhancement}
Our models have shown that although GJ 436b is not enhanced in helium relative to hydrogen, other smaller planets may be. Non-solar ratios of hydrogen and helium could have multiple observable consequences for exoplanet mass, radius, and atmospheric spectral characterization. 

Inferences of planet compositions from transit, transit timing variation, and radial velocity observations rely on comparing the observational constraints on the planet mass and radius to mass-radius relations computed from planet interior structure and evolution models. As discussed in \S~\ref{sec:struc}, increasing the helium mass fraction $Y$ of a primordial envelope from solar to $Y=0.40$ can decrease modeled planet radii by more than a few percent -- a level exceeding the observational precision of planetary radii in the era of Gaia \citep{StassunEt2017AJ}. This additional dimension of planet compositional diversity has so far been largely neglected in the interpretation of planet mass-radius measurements. Assuming solar ratios of hydrogen to helium in the primordial envelopes of close-orbiting sub--Neptune-size planets would lead to a systematic underestimation of the envelope mass fractions of planets that have been sculpted by atmospheric escape.

A depletion of hydrogen and relative enhancement of helium in sub-Neptune-size planet envelopes could have observable consequences for their atmospheric spectra. Increasing the mean molecular weight and decreasing the proportion of hydrogen relative to helium in a planet's atmosphere will decrease the atmospheric scale height and lead to less pronounced absorption features \citep{MillerRicciEt2009ApJ}. The proportion of hydrogen relative to helium can also affect the atmospheric chemistry and equilibrium molecular abundances (e.g., decreasing the proportions of CH$_4$ relative to CO).  \cite{2015ApJ...807....8H} originally proposed a helium enhanced scenario to explain the lack of methane observed in GJ-436b's emission spectrum. Finally, the recent detection of the theoretically predicted \citep{Seager&Sasselov2000ApJ, Oklopcic&Hirata2018ApJ} forbidden 10830 Angstrom He line \citep[e.g.,][]{SpakeEt2018Nature, AllartEt2018Sci, MansfieldEt2018ApJ} opens the possibility of directly detecting the escape of helium from sub-Neptune planets. 

\subsection{Summary}\label{sec:conc}
We implemented the new MESA routine from \S~\ref{sec:methods} to iterate the stages of planet evolution, including modeling planets with varying initial atmospheric helium mass fractions. We ran models across a wide parameter space of mass, envelope fraction, orbital separation, and composition to observe the effects of coupled thermal and compositional evolution. We evolved models starting from 2.0 - 25.0 $M_{\oplus}$, 0.18 - 0.40 helium fraction, $f_{\rm env}$=0.001 - 0.200, and orbital separations from 0.01 to 1.00 AU. We showed the results of helium enhancement on atmospheric structure, as well as how helium enhancement affects $M_p$ --- $R_p$ relations of sub-Neptune mass planets.

Preferential loss of hydrogen can lead to planetary envelopes that are significantly enriched in helium and metals. Evolving planets with initial solar abundances under the mass loss regime from \citep{2015ApJ...807....8H}, we found planets with helium mass fractions in excess of 0.40 and metal mass fractions in excess of 0.10. In addition to helium enhancement, we found planets that had become remnant cores through envelope erosion. We characterized how initial envelope mass, planet mass, and irradiation flux influence the evolution of planet atmospheres.

Helium enhancement is constrained to highly irradiated planets with small initial envelopes. We showed that GJ 436b-mass planets older than $\sim$ 2.5 Gyr with $R_p\lesssim 3.00~R_\oplus$, $f_{\rm env} < 0.5\%$, and irradiation flux $\sim$10$^1$--10$^3$ times that of Earth are likely candidates to have atmospheres with significant helium enhancement. In light of the abundance of short-period, sub-Neptune mass exoplanets that have been recently discovered, this result further expands on the expected diversity of an already incredibly varied population.

Last, we found GJ 436b to not be a suitable candidate to have undergone helium enhancement. GJ 436b's radius places it outside the regime for which hydrodynamic mass loss produced noticeable changes in atmospheric helium abundance. However, we do find that many GJ 436b-mass planets with smaller initial envelope could become helium enhanced, as shown by Figure~\ref{GJ_Ages}. Future work will expand these results and compare our models with the observed planet populations. Helium enhancement models, along with our interior structure simulations, can be valuable predictive techniques to be used in conjunction with new observational surveys.

\section{Acknowledgements}
We would like to thank Dr. Josiah Schwab and Dr. Robert Farmer for their advice regarding modeling low temperature planet envelopes, as well as Dr. Bill Paxton and the broader MESA community. Additionally, we would like to thank the University of Chicago Research Computing Center for their help in running planet simulations. L.A.R. acknowledges support from NSF grant AST-1615315. 

\bibliography{exoplanets}
\end{document}